\documentclass[conference]{IEEEtran}
\usepackage{cite}
\usepackage{amsmath,amssymb,amsfonts}
\usepackage{algorithmic}
\usepackage{graphicx}
\usepackage{textcomp}
\usepackage{xcolor}
\usepackage{stfloats}
\usepackage{makecell}
\usepackage{url}
\def\BibTeX{{\rm B\kern-.05em{\sc i\kern-.025em b}\kern-.08em
    T\kern-.1667em\lower.7ex\hbox{E}\kern-.125emX}}

\usepackage{subcaption}
    
\newtheorem{thm}{Theorem}
\newtheorem{definition}[thm]{Definition}
    
\begin{document}

\title{From Fads to Classics -- Analyzing Video Game Trend Evolutions through Steam Tags}

\author{\IEEEauthorblockN{Nicolas Grelier}
\IEEEauthorblockA{Pullup Entertainment \\
Paris, France \\
nicolas.grelier@pullupent.com}
\and
\IEEEauthorblockN{Johannes Pfau}
\IEEEauthorblockA{Utrecht University \\
Utrecht, Netherlands \\
j.pfau@uu.nl}
\and
\IEEEauthorblockN{Nicolas Mathieu}
\IEEEauthorblockA{Focus Entertainment\\
Paris, France \\
nicolas.mathieu@focusent.com}
\and
\IEEEauthorblockN{Stéphane Kaufmann}
\IEEEauthorblockA{Pullup Entertainment \\
Paris, France \\
stephane.kaufmann@pullupent.com}
}

\maketitle

\begin{abstract}
The video game industry deals with a fast-paced, competitive and almost unpredictable market. Trends of genres, settings and modalities change on a perpetual basis, studios are often one big hit or miss away from surviving or perishing, and hitting the pulse of the time has become one of the greatest challenges for industrials, investors and other stakeholders. In this work, we aim to support the understanding of video game trends over time based on data-driven analysis, visualization and interpretation of Steam tag evolutions. We confirm underlying groundwork that trends can be categorized in short-lived \textit{fads}, contemporary \textit{fashions}, or stable \textit{classics}, and derived that the surge of a trend averages at about four years in the realm of video games. After using industrial experts to validate our findings, we deliver visualizations, insights and an open approach of deciphering shifts in video game trends.
\end{abstract}


\section{Introduction}
Trends emerge in virtually every kind of industry and academic field, steer their direction and development, and often disappear because of saturation, overexposure, or simply making way for new trends \cite{bae2025mathematical}. Beyond fashion, technology, architecture, artistic or cultural shifts, video games are no exception to this phenomenon~\cite{arsenault2009video,nunes2020recent, young2012our}. 
Popular examples of past trends include the boom of Massively Multiplayer Online Role-Playing Games (MMORPGs, e.g. \textit{World of Warcraft}, \textit{Everquest}, or \textit{Star Wars: The Old Republic}). They brought long-term progression, vast social interaction and community building to video games, peaked roughly between 2004 and 2015, but faded out because of the sustained time investment that players could no longer afford, and the rise of free-to-play MOBAs, Battle Royale and Mobile Games that pulled away large parts of the casual player base~\cite{stetina2016moba}.
While some MMORPGs are still alive, other trends ended more abruptly. One instance would be the branch of Toys-to-life games (e.g. \textit{Skylanders}, \textit{Disney Infinity}, \textit{LEGO Dimensions}, or Nintendo's \textit{Amiibos}), which was a blooming market around 2011-2016, but died because of the inherently high consumer costs for figurines and oversaturation from a number of large franchises~\cite{wgecho2017electric}.
Similarly, the era of Motion-Control Games also only lasted from approximately 2006-2012 (sparked by \textit{Nintendo Wii}, challenged by \textit{XBox Kinect} and \textit{PlayStation Move}), as problems of accuracy and reliability were inevitable~\cite{harris2015why}. From 2006-2010, we experienced a wave of rhythm games (e.g. \textit{Dance Dance Revolution}, \textit{Guitar Hero}, or \textit{Rock Band}), which however were unable to keep up their novelty appeal in successor games, while still having to handle high song licensing fees, providing hardware instruments, and suffering from the movement that couch co-op shifted more and more towards online multiplayer.

On the other hand, some trends arguably last. First-Person Shooters (FPS) have never lost popularity since their first successes (e.g. \textit{Wolfenstein} or \textit{DOOM}), Role-Playing Games (RPGs) have become a staple among new releases, and designing for online multiplayer has become common nowadays. Even beyond those high-level genres, low-level ideas such as character customization, open-world design, difficulty levels or accessibility settings have become very established nowadays~\cite{hunicke2005case, squire2007open, brown2021designing, kim2015sense}, and some current advances picked up appealing factors of previous trends (e.g. Virtual Reality games incorporating motion control and in some sense 3D immersion).

In short, it is relatively reasonable to detect factors for successful trends \textit{in retrospection} -- years after a trend has died or became successful -- and find some explanation for that. Yet, detecting the onset of a trend early, and understanding the potential progression and the risk associated to jump on the bandwagon, can be critical information for industrials and investors when deciding to engage in a new project -- especially as the development time of a video game can take longer than a trend lasts.

To advance the understanding of trends and paradigm shifts in video games, we therefore pursue a data-driven analysis, investigating yearly released games on the largest video game distribution platform of our time (\textit{Steam}), and \textit{tags} associated with them.
These tags can be created and voted for by Steam users for each supported game, and it has been shown that they correlate strongly with important game concepts and descriptors \cite{windleharth2016full, li2020preliminary} (e.g. genre, setting, core mechanics, or game modality). At the time of analysis, there are $450$ Steam tags. Thus, they inherently represent factors that are subject to (and descriptive of) larger trends \cite{grelier23icec}.

We point out that this corresponds to an analysis of supply (games released) instead of demand (purchases or playing habits), as our viewing angle represents the interests of industrials making inferences about the state of developed games.
Based on the historical evolution of those tags, we formulate the following research questions:


\textbf{RQ1}: How can we model and analyze the development of a game tag's trend? 

By this, we refer to the evolution of the proportion of released games that were assigned a certain tag. Only considering the absolute number of games released per year is not sufficient because the global number of game released per year has been strongly increasing through the years~\cite{steamdb}, while the global playtime of player has been stable (and even decreasing) in the last ten years~\cite{newzoo2024}.
As the increase in the number of games does not necessarily correspond to an increase in playtime for games with a given tag, the proportion of games featuring that tag better reflects the players' consumption. 

\textbf{RQ2}: Can Bae et al.'s recent general model of \textit{fad}, \textit{fashion} and \textit{classic} trends \cite{bae2025mathematical} be replicated within the domain of video games?

Bae et al. note that trend categories in different branches can have different lifespans (e.g. fashion versus entertainment media), while the core characteristics are similar. Therefore we further pursue:

\textbf{RQ3}: Can we infer general patterns in the duration of tag trends, such as increases, decreases, or intervals between peaks?



\begin{table*}[h]
\centering
\begin{tabular}{l|l|l|}
\cline{2-3}  
& \textit{General} ({\color{blue} \textbf{--}}) \textbf{increases} compared to \textit{high-priority} ({\color{green} \textbf{--}}) & \textit{General} ({\color{blue} \textbf{--}}) \textbf{decreases} compared to \textit{high-priority} ({\color{green} \textbf{--}})\\
\hline
\multicolumn{1}{|l|}{\begin{tabular}[c]{@{}l@{}}Combined \textit{recent} trend\\ score ({\color{orange} \textbf{--}}) \textbf{remains positive}\end{tabular}} & \begin{tabular}[c]{@{}l@{}}
~\\\textbf{Commoditization:}\\
The feature was once a key selling point,\\ but is now becoming mainstream.\\
While more games include it,\\  it is no longer a distinguishing factor.\end{tabular}  & \begin{tabular}[c]{@{}l@{}}
~\\\textbf{Rising staple:}\\
The feature which was previously minor is \\ now becoming a key selling point.\\ 
More games are adopting it as main feature.\end{tabular} \\ \hline
\multicolumn{1}{|l|}{\begin{tabular}[c]{@{}l@{}}Combined \textit{recent} trend\\ score ({\color{orange} \textbf{--}}) \textbf{remains negative}\end{tabular}} & \begin{tabular}[c]{@{}l@{}}
~\\\textbf{Implicit expectation:}\\
The feature is likely still present in many games,\\but players no longer assign the tag because\\it has become an assumed standard.\end{tabular} & \begin{tabular}[c]{@{}l@{}}
\textbf{Decline:}\\
The feature is becoming increasingly niche,\\ gradually fading from the industry.\end{tabular}
\\ \hline
\end{tabular}
\caption{Interpretation scheme of the proposed trend scores in interaction with each other.}
\label{tab:interpretation}
\end{table*}

Lastly, to ground the efficacy of our approach, it is important to present some examples and verify these with the help of expert evaluation, which we do in Section~\ref{sec:qualitative}.
By developing and evaluating a mathematical model of game tag progression, we contribute to a scientific understanding of video game trends and cultural shifts, which can be equally interesting to industry and academia.
In addition to this written contribution, we are also publishing the source code for the full analysis through this github repository\footnote{\url{https://github.com/JohannesPfau/SteamTrends}}, so that readers can replicate and conduct their own analyses for trends of interest.

\section{Related work}
Steam tags have not only aided potential customers in finding appropriate titles, but multiple scientific approaches built on them as a reliable source of categorization. In most of these, tags are used for mere filtering, without further analysis~\cite{lin2018empirical, petrosino2022panorama, pirker2022virtual}. On the contrary, several recent papers make Steam tags the focus of the study. For instance, Windleharth {\em et al.} provide a classification of Steam tags, in particular identifying a list of all genre tags~\cite{windleharth2016full}. Relatedly, Li and Zhang analyze the relationships between genre tags~\cite{li2020preliminary}. Based on that, Li computed the correlations between Steam tags to identify factors that allow for a more suitable classification of games than simply relying on genre~\cite{li2020towards}.

Grelier and Kaufmann introduced the notion of \textit{priority} of a Steam tag \cite{grelier23icec}. For a given game, tags with high-priority are essential descriptors of the game, while tags with low priority are mere features. Using this notion of priority, they later introduce a clustering algorithm relying on Steam tags for grouping games into groups with similar features, and automatically name those groups~\cite{grelier2024automated}.

Under the right circumstances, trends in games with a certain feature can be closely related to success outcomes, as shared industry knowledge and expectations often leads diverse studios to start developing games following (or implementing) certain features that are promising profitability. There are several works on success factors in video games. For \textit{genre} (which is almost always present as tag information), it has been shown that certain genres have an impact on the reception of games \cite{song2016success}. Furthermore, among other game and company variables, genre can be a factor that predicts the revenue of a game and the survivability of its company~\cite{pfau2022predicting, lundedal2023teams}. However, none of these works take into account how genre, or other gameplay features, may be more or less trendy depending on the year, and therefore a genre that was important for predicting success ten years ago may be not important today - which has severe implications on the informativity of these factors, if the development cycle of a game exceeds the time the trend lasts.

Echoing this, Heidenreich {\em et al.} observe that studies identifying tangible factors of what made games successful are still sparse~\cite{heidenreich2023flawless}. They provide a list of success factors for video games, including having a multiplayer mode, the popularity of the brand, and so on. However, they suggest to investigate other potential success factors specific to video games. In particular, we note that they do not consider genres or gameplay features (with the exception of having a multiplayer mode) as potential success factors. We believe that releasing a game with a certain feature at the right moment of its trend is a potential success factor.

\begin{figure}[ht!]
    \centering
    \includegraphics[width=\linewidth]{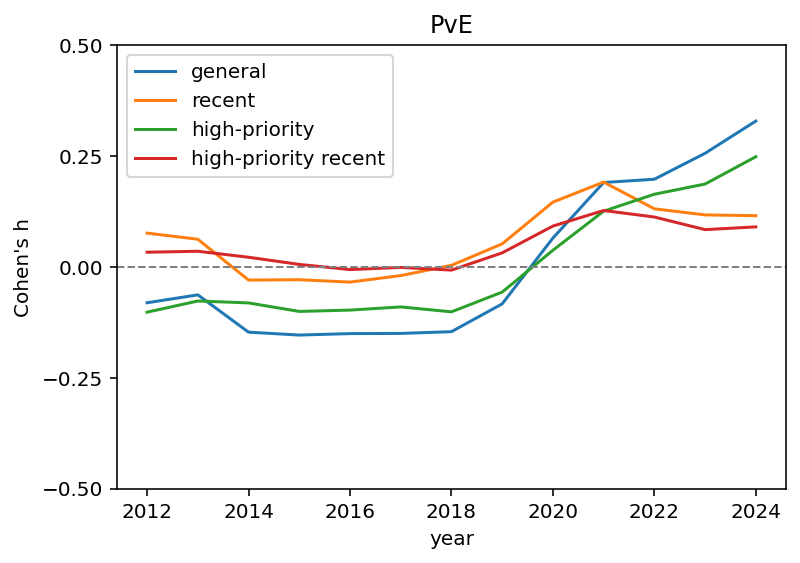}
    \caption{Trend evolution for the tag \textit{PvE}, as measured with different considerations regarding \textit{priority} and \textit{future-dependency} (see Sections \ref{sec:characteristics} and \ref{subsec:four_metrics}).}
    \label{fig:PvE}
\end{figure}

\section{Methodology}
\label{sec:methodology}
Towards developing an analysis suitable for answering \textbf{RQ1}, we first compute four curves that show the evolution of a tag through the years (see Figure~\ref{fig:PvE} for an illustration with the tag \emph{PvE}, which is currently trending), illustrating the global development of a trend with respect to its \textit{priority} and \textit{future-(in)dependency} (as explained below). We chose to analyze data on a yearly scale rather than a more granular one to eliminate seasonal effects, such as the surge in releases at the end of the year.
As four curves might be difficult to interpret, and since the four curves are highly correlated (see Table~\ref{tab:correlation}), we deployed a sparse principal component analysis (PCA)~\cite{zou2006sparse}. The sparse PCA allowed us to combined the two future-independent metrics into one, by doing a weighted average, while preserving most of the relevant information. Depending on the respective behavior of the three remaining curves, the trend evolution can either follow a consistent progression among all curves (general, recent, high-priority), or deviations among these curves can appear. The latter cases are of special interest, as we can infer trend-related patterns using the interpretations of Table \ref{tab:interpretation}.

\subsection*{Dataset-related special characteristics}
\label{sec:characteristics}
While introducing our research questions, we explained that we will focus on the evolution of the proportion of games containing a certain tag. To address our research questions extensively, two additional levels of analysis are helpful to interpret underlying mechanisms of how users assign tags and how tag popularity evolves over time:
    
    \textit{Priority vs Non-priority}: As a game is affiliated with a set of tags, the importance of each single tag should be taken into consideration when interpreting trends. For this, we adapt the proportional priority calculation of related work~\cite{grelier23icec}. This adds a dimension of 
    mainstream/widespread usage (if commonly used among many games, but not essential to a game) versus ``core'' usage (if highly descriptive for a game). In incorporating this dimension, we aim to refine our insights towards \textbf{RQ1}.
    
    \textit{Future-dependency vs Future-obliviousness} (when comparing a proportion to a benchmark): In order to be exhaustive in our description of a trend (\textbf{RQ1}), a crucial point of information is its peak, which must be computed on the whole timeframe of data available. This is however not the only important factor (especially in hindsight to \textbf{RQ2} and \textbf{RQ3}), which necessitate the precise onsets of the rise and decline of a trend. In that case, taking all available data into account may create biases: Let us assume for example that a tag trend increases dramatically during the last three years of the analysis. This surge would elevate the overall average usage of the tag across the entire study period, potentially rendering earlier signals of growth less apparent. However, weak signals of this increase may have emerged prior to the final three years and could be identified by comparing the usage of a given tag in year $Y$ exclusively with its usage in years preceding $Y$. Employing both future-dependent and future-oblivious analyses enables a more comprehensive approach.

To answer \textbf{RQ3}, we first tried to make a list of all duration patterns related to tag trends we could investigate. We tested total trend duration (between the onset and the offset), duration of the trend increase (between the onset and the peak), duration of trend decrease (between the peak and the offset) and duration between two trends for tags with cyclical trend patterns. For all these durations, we manually computed their distributions. 


Taking inspiration from Bae et al.~\cite{bae2025mathematical}, we classify some tag trends into fads, fashions and classics towards answering \textbf{RQ2}. As they suggest, we consider the speed of decrease in trendiness from the peak to the offset to make this classification. Therefore, among the tag trends identified while answering \textbf{RQ3}, we only consider trends that peaked before 2021 in order to have enough data points to identify the decrease speed.


\section{Mathematical Model}



To answer \textbf{RQ1}, we will define four curves, depending on \textit{Priority vs Non-priority} and \textit{Future-dependency vs Future-obliviousness} considerations, while considering the proportion of games with a given tag. However, simply considering the proportions may create false trend signals, coming from actually very small increase in the proportions. Indeed, imagine that a tag $T$ is assigned to $0.001$\% of the games over all previous years, but that this current year it is assigned to $0.01$\% of the games. It seems wrong to say that $T$ is currently trending, when it keeps being assigned to a negligible proportion of games.

For this reason, we use Cohen's h when comparing two proportions~\cite{cohen2013statistical}. For two proportions $p_1$ and $p_2$, Cohen's h is defined as $\phi(p_1)-\phi(p_2)$, where $\phi (p)= 2 \arcsin (\sqrt{p})$. Therefore, Cohen's h ranges from $-\pi$ to $\pi$, and reaches high value only if we have both $p_1$ being larger than $p_2$ and $p_1$ being large enough in itself.

\subsection{Trendiness and Essentiality}
\label{subsec:four_metrics}

Throughout Subsection~\ref{subsec:four_metrics}, we use the following notation: Let us denote by $Y_1, \dots, Y_K$ some $K$ consecutive years, and let $T$ be a tag. We denote by $p_i$, for $1 \leq i \leq K$, the proportion of games released in year $Y_i$ that have the tag $T$. For our current analysis, $Y_1$ is the year $2012$ and $Y_K$ is the year $2024$, but our methodology applies to any year range. Tags were introduced by Steam in $2014$, and then retroactively assigned to games by players. Thus we deemed reasonable to start our study a few years before tags were introduced, but not too early either. While pre-processing the dataset, we noticed that the metrics we introduce are much more noisy before $2012$, which is the reason why we start our analysis at this date.

We introduce a first metric, the \emph{general trend score}, which quantifies the trendiness of a tag in a specific year relative to its trendiness across all considered years. Recall that the number of games released per year has been increasing rapidly. When evaluating the trendiness over all considered years, we prevent recent years from having disproportionate influence.

\begin{definition}[general trend score]
\label{def:general_trend}
     We define $p$ as the average of the $p_i, 1 \leq i \leq K$. The \emph{general trend score} of $T$ in year $Y_i$, denoted by $f(T, Y_i)$, is equal to the Cohen's h of $p_i$ with respect to $p$.
\end{definition}

The general trend score allows one to easily observe the evolution of a tag trend through the years, but one has to keep in mind that the general trend score is future-dependent. As discussed in Section~\ref{sec:methodology}, we introduce a similar but future-oblivious metric:



\begin{definition}[recent trend score]
    For any $5 \leq i \leq K$, we define $p_r(i)$ as the average of the $p_j, i-5 \leq j \leq i$. Thus, $p_r(i)$ is the average of the proportions over the recent years. The \emph{recent trend score} of $T$ in year $Y_i$, denoted by $f_r(T, Y_i)$, is equal to the Cohen's h of $p_i$ with respect to $p_r(i)$.
\end{definition}

The window of \textit{5} years has been chosen for this analysis because of the common time frame of 3-5 years of AAA-game development time \cite{schell2014art,dix2007game}, but can be parameterized at will.
%
%
Both definitions above do not take into account the notion of \textit{priority}~\cite{grelier23icec}. For this reason, we introduce two last metrics, similar to the previous ones, but only considering high-priority tags. We say that a tag assigned to a given game has \emph{high-priority} if its priority is at least $0.6$. This value has been empirically chosen by considering the cumulative histogram of all positive priorities over all tags and games, and selecting a value at the elbow. However, it may also be parametrized at will. We define $p^h_i$, for $1 \leq i \leq K$, the proportion of games released in year $Y_i$ that have the tag $T$ with high priority.

\begin{definition}[high-priority trend score]
     We define $p^h$ as the average of the $p^h_i, 1 \leq i \leq K$. The \emph{high-priority trend score} of $T$ in year $Y_i$, denoted by $f^h(T, Y_i)$, is equal to Cohen's h of $p^h_i$ with respect to $p^h$.
\end{definition}

\begin{definition}[high-priority recent trend score]
    For any $5 \leq i \leq K$, we define $p^h_r(i)$ as the average of the $p^h_j, i-5 \leq j \leq i$. Thus $p^h_r(i)$ is the average of the proportions over the recent years. The \emph{high-priority recent trend score} of $T$ in year $Y_i$, denoted by $f^h_r(T, Y_i)$, is equal to the Cohen's h of $p^h_i$ with respect to $p^h_r(i)$.
\end{definition}

\subsection{Applying a sparse PCA}
When looking at the four curves obtained by the four metrics from Subsection~\ref{subsec:four_metrics}, we observe a strong correlation (see Table~\ref{tab:correlation}). This is not surprising, as we are always considering the Cohen's h of a proportion of games released a given year with respect to other years (the differences coming from considering all tags or only high-priority tags, and considering all years or only the five previous ones).

\begin{table}[h]
\centering
\begin{tabular}{l|l|l|l|l|}
\cline{2-5}
                                           & general & high-priority & recent & \makecell{high-priority \\ recent} \\ \hline
\multicolumn{1}{|l|}{general}              & 1       & 0.81          & 0.69   & 0.54                 \\ \hline
\multicolumn{1}{|l|}{high-priority}        & 0.81    & 1             & 0.59   & 0.68                 \\ \hline
\multicolumn{1}{|l|}{recent}               & 0.69    & 0.59          & 1      & 0.79                 \\ \hline
\multicolumn{1}{|l|}{high-priority recent} & 0.54    & 0.68          & 0.79   & 1                    \\ \hline
\end{tabular}
\caption{Pearson correlations between the four trend scores.}
\label{tab:correlation}
\end{table}

This invites us to condense the analysis towards higher readability, interpretability, and a streamlined metric. The most common approach to dimension reduction is to apply a Principal Component Analysis (PCA)~\cite{jolliffe2016principal}.
We do not center nor normalize our data, as Cohen's h,  which compares a specific data point (a tag-year pair) to a benchmark (a tag over a broader period), inherently acts as a form of centering. Regarding normalization, we apply PCA to inherently correlated signals that exist in different dimensions, with one representing a derivative-like transformation of the other (recent vs. general trends). Thus, we \textit{expect} the weights to reflect the relative importance of derivative and non-derivative components, a distinction that would be lost with normalization.

The principal components of a PCA are generally hard to interpret, which is why we use the sparse PCA algorithm~\cite{zou2006sparse}. This algorithm forces a representation where each principal components is expressed as a weighted sum of a few features. We apply a sparse PCA with three principal components. This provides a way of preserving the variance of the data as best as possible, while having only three curves to look at. The suggested principal components are:
\begin{enumerate}
    \item the global trend score,
    \item the high-priority trend-score,
    \item a weighted average of the recent and high-priority recent trend scores, with coefficients $0.82$ and $0.57$.
\end{enumerate}

We use the coefficients given by the sparse PCA, and define the \emph{combined recent trend score} as $f_c(T, Y_i) := 0.82 f_r(T, Y_i) + 0.57 f^h_r(T, Y_i)$. We now have only three scores to consider to assess the evolution of a tag trend score.

\begin{figure}[ht!]
    \centering
    \includegraphics[width=\linewidth]{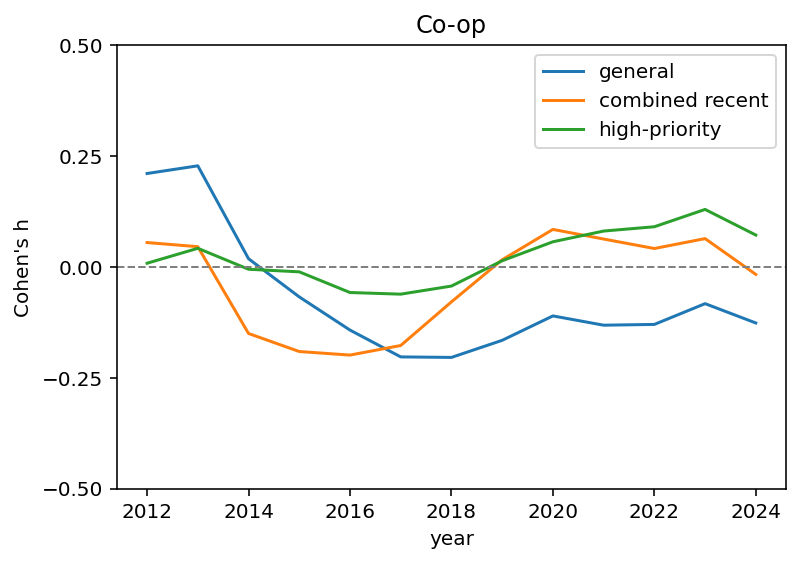}
    \caption{Curves measuring the trend evolution of the tag \emph{Co-op}.}
    \label{fig:co-op}
\end{figure}
\subsection{How to analyse the trend scores}

If the combined recent trend score is far above $0$, then the proportion of games having the considered tag is much higher than the ones from recent years before. This means that the tag is becoming more trendy. See Figure~\ref{fig:PvE} for an example of a tag currently trending\footnote{The combined recent trend score was not represented in Figure~\ref{fig:PvE} for the sake of clarity, but as it is a weighted sum of the red and orange curves, which essentially overlap, the combined recent trend score has the same shape.}.

If the general trend score increases with respect to the high-priority trend score, it means that the tag is becoming more mainstream, more expected by players, and less an essential descriptors of games. On the contrary, if the high-priority trend scores increases compared to the general trend score, the tag is becoming more essential. This can be observed in Figure~\ref{fig:co-op} with the tag \emph{Co-op}.
Again, Table~\ref{tab:interpretation} presents how to interpret the scores when both events happen at the same time.

\subsection{The general duration of a trend increase}
\label{sec:duration_trend_increase}

It often happens that a tag becomes trendy for a few years before declining.
With special regards to \textbf{RQ3}, we here present a method to assess the general duration of a trend increase. 
%

We define \textit{trend increase periods} as consecutive years in which the combined recent trend score remains positive. For each tag, we identify the longest such period, which may be zero if the recent trend score is negative for all years.

Intuitively, the combined recent trend score of a tag $T$ at year $Y_i$, denoted as $f_c(T,Y_i)$ is influenced by its values in previous years, $f_c(T,Y_j)$ for $j<i$. This dependency underlies the concept of trend effects: if $f_c$ has remained positive for several consecutive years, it is likely to turn negative in subsequent years as the tag falls out of fashion. Conversely, in the absence of trend effects, the values $f_c(T,Y_i)$ for a given tag across different years would be independently distributed.

\begin{figure}[ht]
    \centering
    \includegraphics[width=\linewidth]{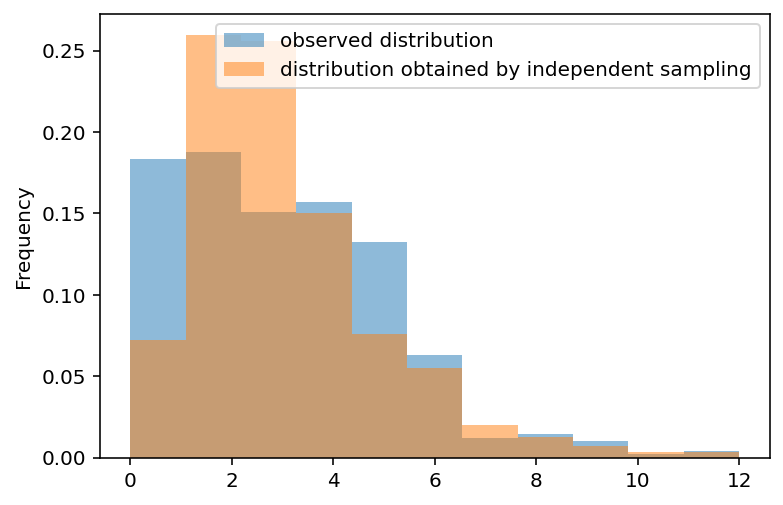}
    \caption{Histograms of maximum duration of trend increases in years. In blue the histogram obtained with the real distribution, in orange the histogram obtained by independent sampling.}
    \label{fig:trend_increase}
\end{figure}

To rigorously analyze the typical duration of trend increase periods, we compare the histogram of these maximum durations to the distribution we would obtain if $f_c$ values were independently sampled. This allows us to contrast the actual observed trend durations with a hypothetical scenario where no trend effects exist. Figure~\ref{fig:trend_increase} depicts these two histograms.

Both histograms exhibit a mode around two years.  However, only the histogram derived from actual data, shown in blue, displays a second peak at four years. As discussed above, this second mode results from trend effects, supporting our claim that trends typically last around four years.

As mentioned in Section~\ref{sec:methodology}, we conducted the same experiment using other durations, including total trend duration, duration of trend decrease and interval between successive peaks. However, there were no strong differences between the histogram obtained with the observed distribution and the one obtained by independent sampling. Since we could not detect trend effects in these cases and due to space constraints, we have omitted these histograms from the paper.

\subsection{Trend Classification}
\label{subsec:trend_classification}
%
Taking inspiration from recent work on general trends by Bae et al.~\cite{bae2025mathematical}, durations for trends might be very complex and dependent on a number of hidden variables, yet certain systematic differences in the development of trends can be observed. Their proposed model is demand-driven, while our method rests on supply-driven methods, therefore we do not apply the very same mathematical model, but draw from their underlying conceptual definitions of \emph{fads} (very short), \emph{fashions} (medium), and \emph{classic} (long) trends.

In the setting of Bae et al.~\cite{bae2025mathematical}, they consider $N$ individuals, who may be interested in adopting a trend, currently adopting a trend, or having left a trend. In our setting, the individuals are actually games. Although a game may evolve with updates, we assume for simplicity that a game either adopts or does not adopt a trend, indicated by the fact of having a tag or not. In our setting, the individuals are not the same through years, but we think of them as cohort of games released at a given year. It is not that individuals stop adopting a trend, but instead that the new cohorts of games do not adopt an older trend.

In their setting, the number of individuals $N$ is constant through time~\cite{bae2025mathematical}. The analogy could work if the number of games released per year was constant too, but it is actually increasing fast~\cite{steamdb}. Thus, their proposed differential system does not apply to our setting. However, we still intent on following their general guidelines:

\begin{itemize}
    \item A trend is a fad if the number of adopters vanishes in finite time,
    \item A trend is a fashion if the number of adopters decreases exponentially fast,
    \item A trend is a classic if the number of adopters decreases polynomially fast.
\end{itemize}

We used all tags with at least four years trend increase as defined in Section~\ref{sec:duration_trend_increase}, bringing it down from $450$ to $194$ tags. In contrast to the former trend duration increase estimation, this analysis necessitates information about the entire trend period ideally. Thus, we further excluded tags with a trend evolution that did not peak four years before conducting the analysis (i.e., in $2021$). 
This left experts with $66$ candidate tags to qualitatively assess, from which we report a selection of examples that were most intensely discussed and illustrated the discussed trend types well.




\section{Qualitative Evaluation}
\label{sec:qualitative}

To evaluate whether the mathematical models proposed in the preceding sections effectively identify trends and analyze their behaviors, we conducted a qualitative assessment with two industry experts, the third and fourth authors, each with decades of experience in game development and market analysis from having worked at Ubisoft and Focus Entertainment. Using semi-structured interviews, we shared the final visualizations along with a basic explanation of our methodology and examined their interpretations. Figure~\ref{fig:PvE} and~\ref{fig:co-op} were presented and we asked the experts whether they aligned with their perceptions of trend evolution.

With respect to whether our metrics are generally representative for the evolution of a tag's trend (and intuitive to interpret), we first refer back to the already illustrated examples of \textit{PvE} (see Figure \ref{fig:PvE}) and \textit{Co-op} (see Figure \ref{fig:co-op}).
The idea that PvE grew trendy (in terms of game releases) after 2018 aligned with the experts' perception of the market, as \textit{``in the early 2010s, successful games with PvE campaigns, such as \textit{Minecraft} (2011), already existed. However, it was the resounding success of \textit{Destiny} (2014) and the high expectations surrounding the release of \textit{The Division} (2016) (despite some player disappointments) that sent a strong signal to the market. These games demonstrated that PvE elements, especially for endgame strategies, could engage players and be successful.''} The consequently happening \textit{``surge in releases''} was \textit{``delayed by a year or two, [...] since PvE games require longer development and balancing''} (on top of the previously mentioned common development time of 3-5 years). \textit{``Subsequently, the consistent release of PvE hits nearly every year, such as Destiny 2 (2017), Monster Hunter: World (2018), Deep Rock Galactic (2020), Valheim (2021), Baldur’s Gate 3 (2023), or HellDivers II (2024), has kept the momentum of PvE-based game development rolling''}, which can be intuitively derived from our visualized metrics.

In the case of \textit{Co-op} as a mechanic, experts underline that this \textit{``has shifted from being an optional mode to a central pillar of game design, explaining its rise as essential''} (or, as per our metrics, \textit{high-priority}). Furthermore, the \textit{``rise of survival games (Ark) or narrative-driven co-op titles (It Takes Two) has emphasized teamwork and shared progression, so that developers now design games around cooperative mechanics, making multiplayer interactions more immersive and engaging''}. Together with this movement, the expectations of the player communities similarly increased, as it was reported that game franchises that contained a Co-op submode (like Assassin's Creed) had to find out that they \textit{``needed to have either a really polished Co-op mode or none (Co-op disappeared after AC Unity)''}, further strengthening the explanatory potential of our \textit{high-priority} metric.


\begin{figure}[ht]
    \centering
    \includegraphics[width=\linewidth]{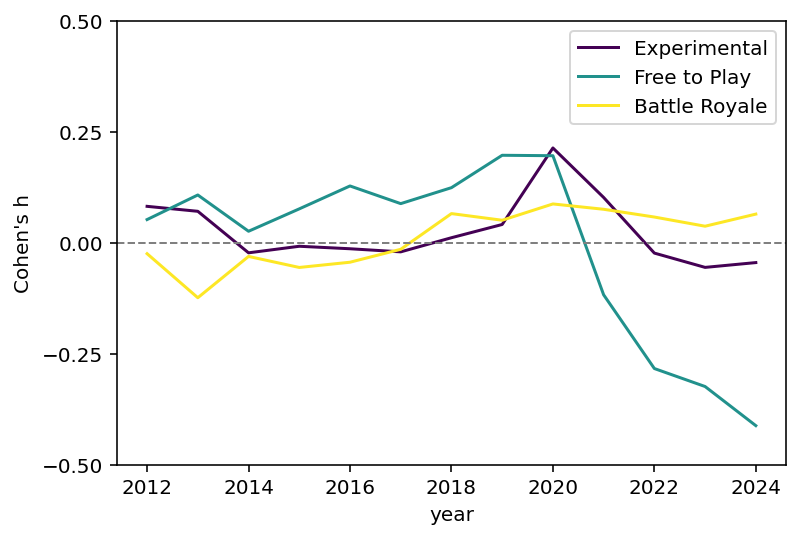}
    \caption{The general trend scores of \emph{Experimental} (fad), \emph{Free to Play} (fashion) and \emph{Battle Royale} (classic).}
    \label{fig:end_trends}
\end{figure}

Regarding \textbf{RQ2}, we presented general trend score curves for three different tags that are conceptually well-defined, which showed a clear trend evolution as indicated by the metric in Section \ref{subsec:trend_classification}, and for which justified elaborations could be derived from industrial knowledge. As depicted in Figure~\ref{fig:end_trends}, this resulted in \textit{Experimental}, \textit{Free to Play} and \textit{Battle Royale}, interpreted as a fad, a fashion and a classic, respectively.

For \textit{Experimental} (e.g. \textit{Townscaper}, \textit{Inscryption}, \textit{Manifold Garden} or \textit{Trombone Champ}), our methodology allowed us to discover a subtantial but short increase in the proportion of experimental games around 2020. This tag was classified as a fad, but explanations were kept careful, as the meaning of \textit{experimental} was perceived as so broad as that there is no easy generalization of this trend.

The abrupt disappearance of \textit{Free to Play} games (fashion) was however well substantiated, as they \textit{``rely on huge retention mechanics (hence continuous development with seasons, events etc.) and microtransactions (hence an always renewed in-game store) to be profitable - this makes them expensive to produce with high risk''}. Furthermore, the market for those became saturated, as \textit{``after the appearance of first hits [...], gigantic games like League of Legends, Fortnite or Valorant took control of the audience, making it utterly difficult for new studios to take the risk to create a new free-to-play''}. While being immensely profitable, it is not a concept new developers should be advised to lean into, as it remains \textit{``a mature market, with no new entrants''}.

After its introduction through games like \textit{H1Z1} (2016) and \textit{Player’s Unknown BattleGrounds} (2017), \textit{Battle Royale} \textit{``really emerged as a genre''}, which has been identified to be a classic trend. Even when excluding the main genre titles with maintained player bases, \textit{``Battle Royale mechanics (shrinking maps, last player standing…)''} were introduced to more and more newly released games as the main driver \textit{``(Fall Guys: Ultimate Knockout, Escape from Tarkov) or temporary battle royale game modes (Rogue Company: Battle Zone, Fallout 76: Nuclear Winter, Rocket League: Knockout)''}.

\begin{figure}
\begin{subfigure}{.5\textwidth}
    \centering
    \includegraphics[width=\linewidth]{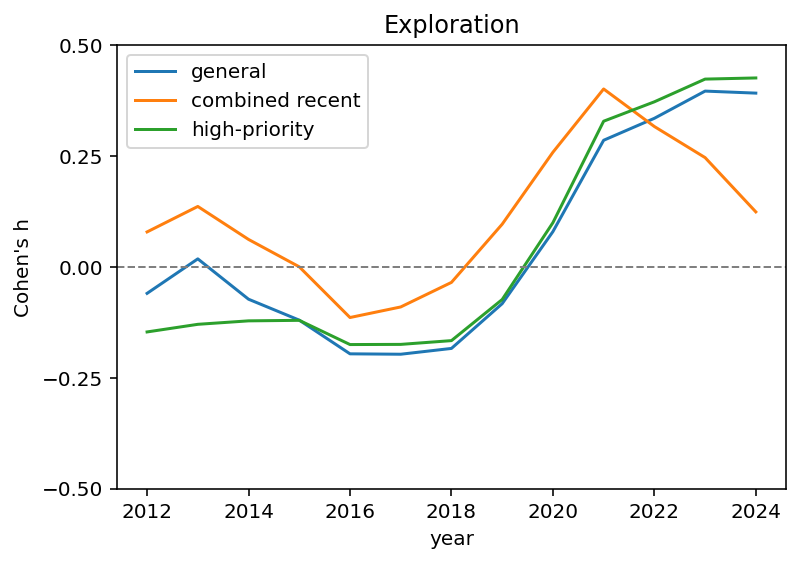}
\end{subfigure}
\begin{subfigure}{.5\textwidth}
   \centering
    \includegraphics[width=\linewidth]{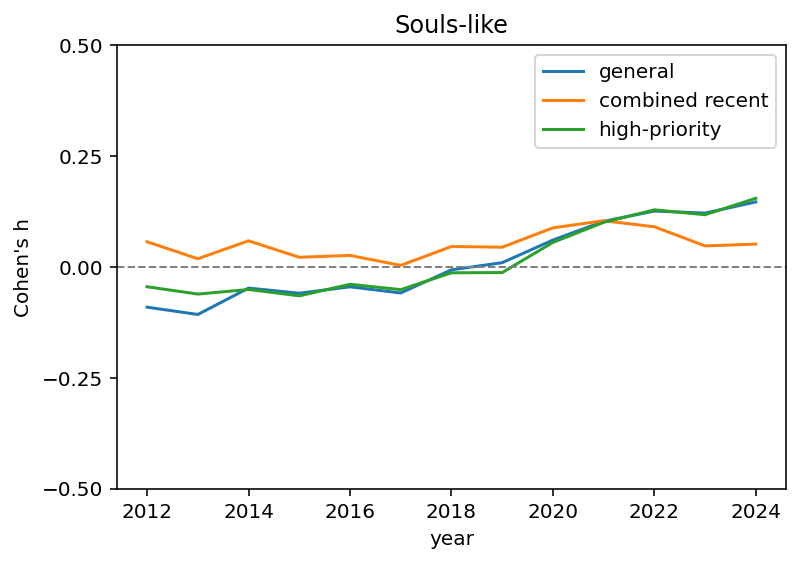}
\end{subfigure}
\caption{The trend curves of \emph{Souls-like} and \emph{Exploration}, two tags that are currently trending.}
\label{fig:current_trend_increase}
\end{figure}

We aim to highlight current notable tag evolutions brought up by our approach in \textbf{RQ3}. Although our reasoning showed that trend increases last four approximately four years, there are of course exceptions, e.g. \textit{Souls-like} and \textit{Exploration} (see Figure~\ref{fig:current_trend_increase}), which we selected to qualitatively analyze. 

The rise of both the general as well as high-priority share of \textit{Souls-like} games correspond to a slow transformation from a niche market into something more mainstream and genre-defining. This is cleanly visualized in Figure \ref{fig:current_trend_increase} (as Cohen's h did not return to its initial value, but stayed high after the trends emerged in 2019), and supported by industrial statements' that they \textit{``managed to become less and less niche through the years by managing to extend its influence outside From Software's games. Striking examples of this spread are Hollow Knight (2017), a Metroidvania game which sold above 20 Millions copies [..., and] encapsulates mechanics from Souls-like games, [or] Rogue-likes like Dead Cells (2017). [...] Elden Ring (2022) also reinvented the Souls mechanics and made them widespread by mixing them with an open-world, increasing the appeal for the genre''}.

As for \textit{Exploration}, this feature gained a lot of momentum after the release of titles like \textit{Breath of the Wild} (2017), \textit{``using open-world in a very Ubisoft-like formula but allowing for more freedom. While Nintendo Switch’s launch hit cannot be taken as a sole reason for the current rise of the trend,''}, experts reported that it \textit{``gave developers a view of [the] audience’s eagerness for [the] feeling of exploration and freedom''}. In addition to genre-defining titles, the importance of exploration as a central or secondary mechanic becomes apparent, as exemplified by games from \textit{``very diverse genre[s] like Firewatch (narrative and linear), Outer Wilds (Knowledge-vania), Subnautica (Survival)  or Hollow Knight (Metroidvania)''} confirming \textit{``the potential of bringing exploration [into] games, increasing its appeal for developing new ones''}.

For broader insight into the outcomes of all tags and custom analyses, readers are invited to consult our open repository¹.


\section{Limitations \& Future Work}

Our study relies solely on Steam tags, which, despite numbering around 450, do not cover every significant feature of video games. Moreover, these tags are user-assigned and may be subject to errors, inconsistencies, or deliberate mislabeling (e.g., \textit{Elden Ring} being tagged as \emph{Family Friendly}). Automatically detecting outliers could help quantify their impact and improve the robustness of our findings.

A key limitation is that we assume a game possesses a feature if it has the corresponding tag. However, since users can assign new tags to older games, we cannot track how perceptions evolve over time. This issue is particularly relevant for live-service games, which frequently change through updates, balance patches, and expansions. Similarly, DLCs can transform a game’s identity (e.g., \textit{Outer Wilds} with its horror-themed DLC \textit{Echoes of the Eye}), yet Steam tags aggregate these changes under the base game. As a result, our analysis may misattribute later content updates to the game's original release. While we assume that drastic gameplay shifts are uncommon, especially since live-service games are more expansive to develop and less frequent, further research could help assess their influence. Besides, we can also not guarantee for the completeness of game features being tagged.

Our analysis does not cover the entirety of tags due to their large number, noise and coverage issues. Developing a comprehensive (interactive) visualization method to represent all tags together could provide a clearer overview of overall trends. Additionally, identifying and highlighting tags with significant evolution over time would offer valuable insights into shifting player interests and industry dynamics.

Figure~\ref{fig:trend_increase} shows that the blue histogram of ``trend duration'' suggests a mix of two distributions: one with mode at one year and the second with a mode at four years. More precise modeling of these distributions could provide a deeper understanding of trend durations, allowing for a more nuanced analysis that includes the standard deviation of tag lifespans.

We focused on identifying trends from the perspective of developers and publishers. A similar analysis from the player’s viewpoint, such as examining playtime by games with a given feature, would be valuable. The main challenge is finding a comparable, accessible database with a large set of user-generated tags, which, to our knowledge, is unique to Steam.

It would be interesting to investigate the origins of trends by analyzing their onset in our data. The question whether trends always emerge from a successful game introducing a new feature, or if they can arise organically without a clear pioneer, is subject to our immediate future work.


\section{Conclusion}
Analyzing, interpreting and understanding video game feature trends is of great importance for developers, publishers, investors and academics. To shed some more light into these complex interactions of this market, we constructed a novel, data-driven method of turning publicly available game publication data into insights about trend evolutions. After quantifying the raw number of game releases and associated tags into multiple metrics of interest, we directly export visualizations that can answer tangible questions, like \textit{``Did a tag gain or lose popularity?''} or \textit{``Did it become a more or less essential descriptor of games?''}. Our findings suggest that the increase of trends in this context typically last about four years. Given that video game development often spans several years, this alone communicates a warning to the industry that pursuing trends may not be a viable long-term strategy, as a game's release may coincide with the decline of the trend. This might change if a trend would turn into a classic, for which early detection methods are still to be established.
We conclude our work with a qualitative analysis of field experts that supports our data-driven conclusions, reinforcing the validity of our approach. These insights offer valuable guidance not only for developers and publishers but also for academics studying industry trends, game design evolution, and market dynamics.

\bibliographystyle{IEEEtran}
\bibliography{bib}

\end{document}